# Investigating Collaboration Within Online Communities: Software Development Vs. Artistic Creation

**Giuseppe Iaffaldano**
University of Bari
Bari, BA, Italy
Giuseppe.iaffaldano@uniba.it

## Abstract
Online creative communities have been able to develop large, open source software (OSS) projects like Linux and Firefox throughout the successful collaborations carried out over the Internet. These communities have also expanded to creative arts domains such as animation, video games, and music. Despite their growing popularity, the factors that lead to successful collaborations in these communities are not entirely understood. In the following, I describe my PhD research project aimed at improving communication, collaboration, and retention in creative arts communities, starting from the experience gained from the literature about OSS communities.

## Author Keywords
Online communities; creative collaboration; music composition; overdub; remix; SNA.

## ACM Classification Keywords
H.5.3. Information interfaces and presentation (e.g., HCI): Group and Organizational Interfaces – *collaborative computing, computer-supported cooperative work.*

## Introduction
Online creative communities are virtual groups whose members volunteer to collaborate over the Internet to produce software, music, movies, games, and other cultural products. Despite the growing popularity of online communities, the factors that lead to success are not entirely understood [9]. For instance, we ignore whether success factors are domain dependent [6]. Previous research has established that open-source software (OSS) development is a form of creative peer-production [7] and that successful collaborations between developers in OSS communities depend on both social and technical factors [11]. Likewise, Luther et al. [6] found that participants' social reputation is key to the successful completion of collaborative animation efforts in online arts communities. Other studies suggest that active feedback actions, such as commenting, are fundamental to the success of creative communities [8]. More recent studies found that users who collaborate online are motivated by their will to learn, socialize, improve their reputation, and gain new opportunities for personal success [4,10]. Thus, research shows that success factors in online creative collaborations are closely related to the social and personal domain [11].

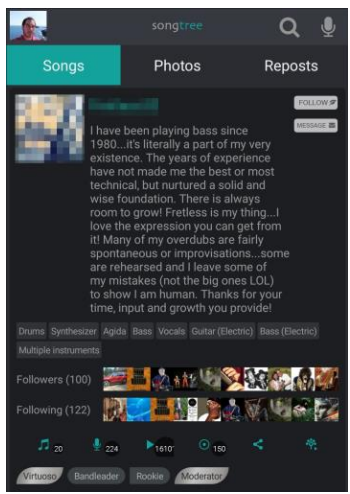

**Figure 1:** The stats available in the profile page of an artist in Songtree.

As part of my PhD research on online creative communities, I am currently investigating Songtree, a community for collaborative music creation [2,3]. The goal is to further the understanding of the factors influencing user activities, participation, and collaboration in such creative communities and whether these factors transfer across domains. I expect to advance the state of the art in the fields of CSCW and HCI by empirically identify the success factors of creative online communities, both specific to the artistic domain and in common with OSS communities; define new guidelines and tools to support collaboration in online communities, specifically with synchronous creative activities (e.g. brainstorming, real time creative collaborations); define guidelines and tools to foster creativity in collaboration within software development communities, where newcomers typically encounter barriers that prevent them from participating.

## Research Questions
*RQ1: What are the success factors of collaboration in online creative communities?*

First, I aim to identify the several success factors that determine the success of the interactions in different types of online creative communities. These factors will be retrieved through the review of the existing literature on collaboration in online communities and tested through empirical studies.

*RQ2: Which of the identified success factors depend on the domain of collaboration?*

Besides, I intend to determine which of the findings from previous research on OSS communities would carry over to the domain of creative arts communities.

## Work and Findings to Date
Songtree is an online creative music community where artists collaborate to the creation of musical tracks. Several social-networking features are available, and users can grow in popularity through their songwriting activity within the community (see **Figure 1**). Songtree allows any user to derive (namely, *overdub*) any song shared in the community leveraging the metaphor of a growing tree (see **Figure 2**). Songtree's overdubbing feature shares many similarities with the Pull Request-based software development model supported by modern code-hosting platforms [2,5]. Considering the existing analogies between the two platforms, my work so far has focused on seeking confirmation that existing findings from prior work in OSS communities would also transfer to artistic communities. More specifically, I have focused on *mining roles and communication* in the Songtree community and *determining the success factors* of the collaborations carried out therein.

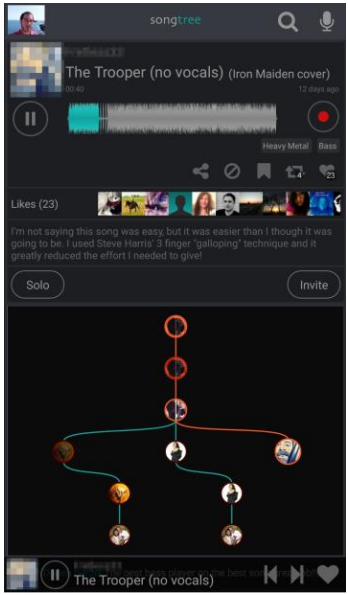

**Figure 2:** The tree of a song (topmost node) with branches generated by the overdubs received.

*Roles and Communication: A Social Network Analysis*
Bird et al. [1] built the social network of the Apache software foundation developers based on their communication through the mailing list. Their goal was to understand the properties of the social network, the differences in role between developers and non-developers, if development and communication activities are correlated, and the effect of developers' status in the project.

In [3], I reported on a study that mapped the research questions of Bird et al. [1] to Songtree in order to identify the main roles, activities, and key members in the community, thus uncovering the underlying communication network. I built a *feedback network,* a directed graph where nodes represent authors and links are comments on songs. Users commenting activity shows a power-law distribution, typical of online communities, where a small group of users is very active in commenting songs whereas the majority of members only sends a few comments. This evidence suggests authors of popular songs receive more comments. A moderate positive relation was also found between the number of comments sent by authors and their related in-degree (i.e., how many commented on an author's songs), suggesting that the commenting activity may contribute to increasing their visibility. As for the roles in the feedback network, we sought differences between the commenting behavior of authors and non-authors. Results showed that a higher percentage of authors do not provide comments, indicating that non-authors are more inclined to giving feedback. Finally, we analyzed the relationship between the commenting and songwriting activities by running a correlation analysis restricted to authors only. Results showed a weak rank correlation between the number of comments sent by a Songtree authors and the number of songs they recorded.

*Antecedents of Overdubs: An Analysis of Success Factors*
Tsay et al. [11] have studied collaboration success factors of in OSS communities. Specifically, they built a logistic regression model and showed that social factors (e.g., developer's status in the project) are more influent than the technical ones (e.g., presence of test cases in a pull request) in taking the decision of accepting or not a pull request in GitHub.

Similarly, I have investigated in [2] which song- and author-related factors can influence the probability of a song to be overdubbed. The results showed that the overdub likelihood of a song is strongly and positively associated with the amount of reactions (e.g., likes, bookmarks) that it generates. Some song-related variables, instead, were found to be negatively associated with the likelihood of a song to be overdubbed (e.g., time since the upload). As for the author-related predictors, author ranking in Songtree and the presence of the author profile picture were found to be positively associated with the overdub likelihood.

## Next Steps

*Future work on Social Network Analysis*
I intend to analyze: (i) collaborations in song trees to uncover cliques of users forming distributed virtual bands and their motivations; (ii) the differences in role and behavior between *core* and *peripheral* community members.

*Future Work on Success Factors*
I intend to improve the statistical model by: (i) including new predictors of successful collaboration using SNA measures; (ii) conducting a longitudinal study to study retention and loyalty of community members over time using a survival analysis approach.